
\input phyzzx.tex

\Pubnum={\vbox{\hbox{QMW 92/2}\hbox{Imperial TP/91-92/19}}}
\pubtype={Revised Version}
\date={April 1992}

\titlepage

\title{The canonical structure of  Wess-Zumino-Witten models}

\author{G. Papadopoulos}

\address{Department of Physics\break Queen Mary and Westfield College\break
              Mile End Road\break London E1 4NS, UK.}
\andauthor {B. Spence}
\address{Blackett Laboratory\break
Imperial College\break London SW7 2BZ, UK.}

\abstract{The phase space of the Wess-Zumino-Witten model on a circle with
target space a compact, connected, semisimple Lie group $G$ is defined and the
corresponding  symplectic form is given.    We present a careful derivation
of the Poisson brackets of the Wess-Zumino-Witten model.  We also study the
canonical structure of the supersymmetric and the gauged Wess-Zumino-Witten
models.}

\endpage

\pagenumber=1


\def \a {\alpha}
\def \d {\partial}
\def \t {\theta}


\REF\wit {E. Witten, Commun. Math. Phys. {\underbar {92} }(1984), 455.}
\REF\blo {B. Blok, Phys. Lett. \underbar{233B} (1989), 359.}
\REF\ale {A. Alekseev and S. Shatashvili, Commun. Math. Phys. \underbar
{128} (1990), 197; Commun. Math. Phys. \underbar{133} (1990), 353.}
\REF\fad {L. D. Faddeev, Commun. Math. Phys. \underbar{132} (1990), 131.}
\REF\bal {J. Balog, L. Dabrowski and L. Feher, Phys. Lett. \underbar
{244B}(1990), 227.}
\REF\fel {G. Felder, K. Gawedzki and A Kupiainen, Nucl. Phys. \underbar {B299}
(1988), 355; Commun. Math. Phys. \underbar{117} (1988),127.}
\REF\gaw {K. Gawedzki, \lq\lq Classical Origin of the Quantum Group Symmetries
in the Wess-Zumino-Witten Conformal Field Theory", IHES/P/90/92.}
\REF\chu {M. F. Chu, P. Goddard, I. Halliday, D. Olive and A. Schwimmer, Phys.
Lett. \underbar {B266} (1991), 71.}
\REF\stern { G. Bimonte, P. Salomonson, A.
Simoni and A. Stern,\lq\lq Poisson Bracket Algebra for Chiral Group Elements in
WZNW Model",  UAHEP 9114 preprint (1992).}
\REF\zuck {E. Witten, Nucl. Phys. \underbar{B276} (1986), 291; G. Zuckerman, in
\lq\lq Mathematical Aspects of String Theory", ed. S.T. Yau, World Scientific,
Singapore (1987); C. Crnkovic and E. Witten, in \lq\lq Three Hundred Years of
Gravitation", ed. S.W. Hawking and W. Israel, C.U.P., Cambridge (1987), 676.}
\REF\osb {C.M. Hull and B. Spence, Phys. Lett. \underbar {232B} (1989), 204;
I. Jack, D.R. Jones, N. Mohammedi and H. Osborn, Nucl. Phys. \underbar{B332}
(1990), 359.}
\REF\ed {E. Witten, \lq\lq On Holomorphic Factorization of WZW and Coset
Models", IASSNS-91/25 preprint.}
\REF\papa {G.Papadopoulos, Class. Quantum
Grav.\underbar{8} (1991), 1311. }
\REF\bab {O. Babelon, Phys. Lett. \underbar
{215B} (1988), 523. O. Babelon and C. M. Viallet, \lq\lq Integrable models,
Yang-Baxter Equation and Quantum Groups", SISSA preprint (1989).}
\REF\isham {C. Isham, \lq\lq Topological and Global Aspects of Quantum Theory",
in \lq\lq Relativity Groups and Topology II", Ed. DeWitt and Stora, (1983) Les
Houches, North-Holland.}


\section {Introduction}
	The  Wess-Zumino-Witten (WZW) models  are two-dimensional
sigma
models with
target space a group manifold and they constitute an important class of
conformal field theories.  In ref. [\wit ] the left and right currents  were
used to study the quantisation of WZW models.   The quantisation of the
Poisson bracket (PB) algebra of the currents leads to two copies of a Kac-Moody
algebra, one copy for each chiral sector. The Hilbert space of the theory is
constructed from the representation theory of these Kac-Moody algebras.

Recently, in refs. [\blo-\stern] the  quantisation of the WZW model was
investigated in terms of the group elements of a loop group.   The quantisation
of the PB algebra of the WZW model in terms of these variables leads to a
quantum group structure characteristic of these and other integrable
two-dimensional models.

The calculation of the PBs of the group elements of the loop group is an
interesting problem. One difficulty in calculating the PBs of the WZW
models from their symplectic form arises from the fact that the phase space
of the theory  is an infinite dimensional space and the  symplectic form
is an operator.  Thus the methods usually applied to finite dimensional
systems are not  directly applicable in this case.  Several authors
[\gaw, \chu, \stern ] attacked
the problem of calculating the PBs of the WZW model
using as a  starting point the definition of the phase space of the theory as
the space of solutions of the model [\zuck ], and as a symplectic form the
one derived from the classical action of the system.  In particular this
approach was studied in detail by the authors of reference [\chu].

In this paper, we give a new derivation of the PBs for the WZW
models.  Our derivation follows from  a
careful study of the structure of the phase space of the WZW models, the
use of sigma model methods and an application of methods in operator theory.
The advantages of our derivation are that the PBs of the WZW model are
constructed from first principles and in a geometric way.   The Jacobi
identities for these PBs are explicitly shown to be satisfied.    These PBs are
consistent with the topology and geometry of the spacetime, which is taken to
be a cylinder $S^1\times \bf{R}$.

To achieve this, we introduce a  model which is related to
the WZW model but which has different left and right monodromies; we call
this the LR model. The phase space $P_{LR}$ of the LR model is an
enhanced version of the phase space $P$ of the WZW model.  The PBs of the LR
model are calculated and it is shown that this theory factorises into  left
and right moving sectors, i.e. the PB of any left sector variable
with any right sector variable
vanishes.  We show how our results agree with those in
the \lq\lq chiral approach"  of ref.[\chu]. The phase space of the WZW model is
shown to be given by  the phase space of the LR model with the addition of a
first class constraint that enforces the equality of the left and right
monodromies.  We also study the canonical structure of the $(1,1)$
supersymmetric and gauged WZW models.

In section two, we outline the WZW model, state its
symplectic
form and set up our notation.  In section three, we derive the PBs of
the LR model and describe the WZW model in terms of the
phase space $P_{LR}$ of this model.  In sections four and five, we examine the
supersymmetric and gauged WZW models.  Finally, in section six we present our
conclusions and comment on the quantisation of the WZW model.

\section { The WZW model}

The WZW model is a sigma model with  Wess-Zumino term whose target space is a
group manifold $G$, where  $G$ is a compact, connected, semisimple Lie group.
The metric of
the WZW model is a bi-invariant metric $h$ on $G$ and the  Wess-Zumino term
is a bi-invariant (closed) three form $H$.  After an appropriate normalisation
of the metric and Wess-Zumino terms the Lagrangian of the WZW model [\wit] is
$$ L= -{k\over 16 \pi}\  \Big(h_{ij}\  \eta^{\mu \nu}\  \partial_{\mu}\phi^i{}
\partial_{\nu}\phi^j +
\  b_{ij}\  \epsilon^{\mu \nu}\  \partial_{\mu}\phi^i\
\partial_{\nu}\phi^j\Big),					\eqn\aone$$
where $\phi$ is a map from a cylinder $S^1\times \bf{R}$ with co-ordinates
$\{(x,t),0\leq x\leq l, -\infty < t< \infty \}$ to the group $G$,
$H={3\over 2}db$, $\eta$ is a metric on $S^1\times \bf{R}$,  $\epsilon$ is a
two form on $S^1\times \bf{R}$ ($\epsilon^{01}=1$), and $i,j=1,\cdots, dimG$.
The metric $h$ and the form $H$ can be expressed as
$$h_{ij}= L^a_i\  L^b_j\  \delta_{ab}= R^a_i\  R^b_j\  \delta_{ab}\eqn\atwo$$
and
$$H_{ijk}={1\over 2} L^a_i\  L^b_j\  L^c_k\  f_{abc} = -{1\over 2} R^a_i\
R^b_j\  R^c_k\  f_{abc} 				\eqn\athree$$
respectively.  $L$ ($R$) is the left (right) frame on the
group $G$, ${f^a}_{bc}$ are the structure constants of $LieG$ and
$a,b,c=1,\cdots, dim LieG$ are Lie algebra indices.

The metric and three form $H$ can be rewritten as
$$\eqalign {h_{ij}&= tr ( g^{-1} \partial_i g\  g^{-1} \partial_jg),
\cr      H_{ijk}&=- \ tr( g^{-1} \partial_{[i}g\   g^{-1} \partial_jg\
g^{-1} \partial_{k]} g).}				\eqn\afour$$
For simplicity,  by $g$ we mean the group element $g\in G$ in some
unitary representation $\pi$ of $G$.   $tr$ is normalised such that $tr(t_a
t_b)=-\delta_{ab}$ where $t$ ($[t_a,t_b]=i\  {f_{ab}}^c\  t_c$) is the
representation of $LieG$ that corresponds to the representation $\pi$ of $G$,
and $g\in G$, $g^{-1}\partial_i g= i\  L^a_i\  t_a$ and $
\partial_i g\  g^{-1}= - i\  R^a_i\  t_a$.

Due to the relative normalisation of the two terms in the
Lagrangian \aone,  the equations of motion are
$$\partial_-(\partial_+  g\ g^{-1})=0,\quad \partial_+ (g^{-1}
\partial_- g)=0 					\eqn\afive$$
where $x^{\pm}=t\pm x$.
The action \aone\ is invariant under the semilocal transformations
$g\ \rightarrow l(x^-)\  g \  r^{-1}(x^+)$
and the corresponding currents are
$$
J_-(x^-) = {ik\over 4 \pi}\  g^{-1}\ \partial_-g,\quad
J_+(x^+) = -{ik\over 4 \pi}\  \partial_+g\ g^{-1}.		\eqn\aseven$$

The general solution of the equations of motion \afive\ can be expressed in
the form
  $$g(x,t) = U(x^+) M^{{2t\over l}}\  V(x^-).	\eqn\aeight$$
where $g, U, V $ are  periodic functions of $x$ and $M$ is on a maximal torus
of the group $G$.
This solution is invariant under  two sets of transformations,
 one  given by the action of the  Weyl group, and the
other being
$$U\rightarrow U\  h,\quad V\rightarrow h^{-1}\  V, \quad
		M\rightarrow M, 			\eqn\aten$$
where $h$ is any element of a maximal torus of $G$.

The phase space
$P$ of the WZW model is the space of functions $\{U, V, M\}$
 up to the relations given by  eqn. \aten
\foot { More precisely, the
phase space of the WZW model is the space of orbits of the group action \aten\
on the bundle space of a fibre bundle with base space a maximal torus of the
group $G$ which parametrises the monodromy and fibre $LG\times LG$ where $LG$
is the loop group of $G$.}.

\section {The Poisson Brackets}

The symplectic form $\omega$ of the WZW model is the space integral of the time
component of the symplectic current $S$ where
$S^{\mu}= \delta \phi^i \wedge \delta ({\d L / \d \partial_{\mu}
\phi^i})$.
The symplectic form in the basis $\{U, V, M\}$ is
$$\eqalign {
\omega =& - {k\over 8 \pi}\bigg[ \int^{\l}_0 dx\ tr\Big(U^{-1} \delta
U\ \partial_x (U^{-1} \delta U) + {2 i\over \l}(U^{-1} \delta U)^2\  A
\cr &
 - \delta V V^{-1}\  \partial_x (\delta V V^{-1}) - {2 i \over \l} (\delta V
V^{-1})^2\  A \Big)
 \cr &
+{2\over \l} tr\Big(
 (U^{-1} \delta U)_0\  \delta M M^{-1} + {2\over \l} (\delta V
V^{-1})_0\  \delta M M^{-1}\Big)\bigg].}           \eqn\btwo$$
where $M=exp i A$ and $(U^{-1} \delta U)_0$ ($(\delta V V^{-1})_0$) is the
constant mode of $U^{-1} \delta U$ ($\delta V V^{-1}$).  $\omega$ is closed,
and independent of the time $t$ and the origin $x_0$ of the interval
$[x_0, x_0+l]$ (we take $t=x_0=0$).
$\omega$ is degenerate along the directions of the transformations \aten.

To continue, we introduce a new model called the LR model.  The phase
space $P_{LR}$ of this model is the space of functions $\{U,V, A_L,
A_R\},$   with symplectic form $\Omega = \omega_L + \omega_R$
where
$$\eqalign{
\omega_L(U,A_L) =&- {k\over 8 \pi}\big{[} \int^{\l}_0 dx\ \{ tr(U^{-1} \delta
U\ \partial_x (U^{-1} \delta U) + {2{}i\over \l}(U^{-1} \delta U)^2\  A_L)\}
\cr &
+ {2\over
\l}\  tr (U^{-1} \delta U)_0\  \delta M_L M_L^{-1}\big{]} }    \eqn\bfour$$
and $\omega_R(V,A_R):=-\omega_L(V^{-1}, A_R)$.

The forms $\omega_L,\  \omega_R$ are  closed provided the monodromies $M_L,\
M_R$ are restricted to be in a maximal torus of $G$ [\gaw], i.e $A_L,\ A_R$ are
in a Cartan
subalgebra of $LieG$.  The symplectic forms $\Omega$, $\omega_L$ and
$\omega_R$ are not degenerate.  The transformations \aten\ do not leave
these symplectic forms invariant.  In particular $\Omega$ is not invariant
because the monodromy of the left sector is different from that of the right
sector.  The Hamiltonian of the LR model is the one given by the  sum of
the Sugawara Hamiltonians of the left and right chiral sectors.

It is clear that the phase space $P_{LR}$ and the symplectic form $\Omega$ of
the LR model factorise, i.e. $P_{LR}=P_L \times P_R$ and $\Omega =\omega_L
\oplus \omega_R$.  In the following we study the inverse of  $\omega_L$ only
since the methods used for this inverse can be  applied straightforwardly to
calculate the inverse of $\omega_R$ and hence the PBs of the theory.

The symplectic form $\omega_L$ can be expressed in a representation independent
way.  Indeed, using $ U^{-1}\delta U =\  i\  t_a\  L^a_i\  \delta X^i$, we can
rewrite it as
$$\eqalign {
\omega_L =&- {k\over 8 \pi}\big{[} \int^{\l}_0 dx\   \{ L^a_i\
\delta X^i\ (\delta_{ab}  \partial_x  - {1\over l}  A^c_L\  f_{cab})\
  (L^b_j\   \delta X^j)\}
\cr &
+{2\over \l}\  \delta_{ab}\ (L^a_i \delta X^i)_0\  \delta
A^b_L\big{]}. }					\eqn\bfive$$

To find the PBs of the left sector of the theory, we study the inverse of
the operator
$$D_{ab}= {d \over dx}{} \delta_{ab} -{1\over l} A^c_L{} f_{cab}.
\eqn\bsix$$ The operator $D$ acts on periodic functions with period $\l$, i.e
it is an operator on a circle of circumference $\l$.  The construction of the
inverse of $D$ is considerably simplified by observing that $A_L$ is a vector
in the Cartan subalgebra $\bf {h}$ of  $LieG$.  In a Cartan-Weyl basis for
$LieG$, the operator $D$ can be rewritten as follows
$$ D_{rs} = {d \over dx}{} \delta_{rs},\quad r,s=1,\ldots,d=dim\bf
{h},							\eqn\bseven$$
$$D_{\a{}-\a }={d \over dx}{} \delta_{\a{}-\a } + {i\over l} \a^r (A_L)_r
							\eqn\beight$$
and
$$D_{-\a{}\a }={d \over dx} \delta_{-\a{}\a } - {i\over l} \a^r (A_L)_r
							\eqn\bnine$$
where $\{\a \}$ is  the set of positive roots of $LieG$.
To invert $D$ we invert each component separately.  $D_{rs}$ is
an elliptic anti-Hermitian operator with a space of zero modes of
dimension equal to the rank of the group $G$ (there is a natural inner
product for the operator $D$ on the circle).  Moreover the operator
$D_{rs}$ is invariant under $O(d)$ rotations.  This operator does not
have a unique inverse. The kernel $K_{rs}(x,y)$ of the inverse operator  \foot
{The kernel K(x,y)
of the inverse  operator  of an operator $D$ on a space $M$
obeys the condition  $ {1\over Vol(M)}\ \int_M
  dy K(x,y)$$ \varphi_{\kappa}(y)$$={\varphi_{\kappa}(x) \over
\kappa}$ where $\kappa$ are the eigenvalues and $\varphi_{\kappa}$ are the
eigenfunctions of the operator $D$ ($\kappa \ne 0$).}
of  $D_{rs}$ is specified
up to a constant matrix.  The antisymmetry of the PBs requires that the kernel
$K_{rs}(x,y)$ be antisymmetric, i.e
$$ K_{rs}(x,y)= - K_{sr}(y,x).		 \eqn\bten$$
The kernels that
obey this condition are
$$ K_{rs}(x,y)= - \Big[(x-y) mod\  \l-{\l \over 2}\Big]
\delta_{rs} + C_{rs}; \quad C_{rs}=-C_{sr}			\eqn\beleven$$
where $C=C(A_L)$ is a constant matrix and $0\le (x-y) mod\ l<\ l$.
  The Jacobi identities of the PBs impose
additional restrictions on the matrix $C$, in fact it is found that $C$ is
a closed two-form on a maximal torus of the group $G$.

There is a unique inverse of the operator $D_{rs}$ which is invariant
under $O(d)$ rotations.
This inverse operator has kernel $K_{rs}$ given
in eqn. \beleven\ with $C=0$.  The zero modes of the integral
operator given by  the kernel $K_{rs}$ ($C=0$) are the constant functions on
the circle, i.e. they are the same as the zero modes of the operator $D_{rs}$.
Therefore $K_{rs}$ ($C=0$) is the appropriate kernel which inverts the
symplectic form outside the zero modes of $D_{rs}$.  Finally, the zero modes of
$D_{rs}$ are \lq\lq conjugate'' to the monodromy $A_L$ in the symplectic form
$\omega_L$ (see the last term of the symplectic form $\omega_L$), giving a
non-degenerate symplectic form and consistent PBs \foot {Note that step
functions are not the kernels of the inverse of $D_{rs}$ on the circle.  One
reason is that they have an even number of discontinuous points when they are
considered on the circle.}.

Next we turn our attention to the operator $D_{\a{}-\a}$.  If $\a^r
(A_L)_r\ne 2\pi n$, $n$ an integer, the operator $D_{\a{}-\a}$ has no zero
modes  and it admits a unique inverse.
The kernel of the inverse is
$$K_{\a{}-\a}(x,y) = {-i {\l \over 2} \over sin({\a A_L\over 2})}
exp\bigg[-{i\a
A_L \over \l} \Big((x-y){} mod\  \l- {\l \over 2}\Big)\bigg].
                                              \eqn\btwelve$$
Similarly the kernel of the inverse of the operator $D_{-\a{}\a}$ is
$$K_{-\a{}\a}(x,y) = {i {\l \over 2} \over sin({\a A_L\over 2})}
exp\bigg[ {i\a A_L
\over \l} \Big((x-y){} mod\  \l- {\l \over 2}\Big)\bigg].
                \eqn\bthirteen$$
The
PBs expressed in the basis given by the maps $X$ are thus
$$\eqalign {
\{X^i(x), &X^j(y)\}_{PB} ={1\over \beta} L^i_a(X(x)){} K^{ab}(x,y){}
L^j_b(X(y))
\cr &
= {1\over \beta}
\big {[}L^i_r (X(x)) \{- ((x-y) mod\  \l-{\l \over 2})\} \delta
^{rs} L^j_s(X(y))
\cr &
+\sum_{\a } {i {\l \over 2} \over
sin({\a A_L\over 2})}\  L^i_{\a}(X(x))   exp\{ {i\a A_L \over
\l} ((x-y){} mod\  \l- {\l \over 2})\}\  L^i_{-\a} (X(y))
\big{]}}						\eqn\bfourteen$$
$$\{X^i(x),( A_L)_r\}_{PB} = - {\l \over \beta} \  L^i_r \eqn\bfifteen$$
and
$$\{( A_L)_r,( A_L)_s\}_{PB}=0,				\eqn\bsixteen$$
where $\beta = - {k \over 8 \pi}$ and $\sum_{\a }$ is the sum over all roots.

These  PBs satisfy the Jacobi identities.  The only
non-trivial Jacobi identity is the vanishing of the expression
$$\{ \{X^i(x),X^j(y)\}_{PB}, X^k(z)\}_{PB} + cyclic.   \eqn\bseventeen$$
To verify this Jacobi identity, we substitute eqn. \bfourteen\ into the
above expression, giving
 $$\eqalign {
{1\over \beta^2}\ \bigg[ l\ &\d_c  K_{ab}(x,y) +l \d_b K_{ca}(z,x) +l\ \d_a
K_{bc}(y,z)  \cr &
- {f_b}^{ed} K_{ad}(x,y) K_{ec}(y,z)
- {f_a}^{ed} K_{cd}(z,x) K_{eb}(x,y)
\cr &
- {f_c}^{ed}  K_{bd}(y,z) K_{ea}(z,x) \bigg]\  L^a_i(x)\  L^b_j(y)\
L^c_k(z), }						\eqn\beighteen$$
where $\d_a ={\d \over \d A_L^a}$.
This  can be shown to be identically zero by substituting the $K$ of
eqns. \btwelve\ - \bfourteen\ into  \beighteen\ and by restricting $C$ to be a
closed two-form, i.e. $\partial_{[r}C_{st]}=0$.

Having calculated the PBs in the $(X, A_L)$ basis, we can calculate the PBs of
any other functionals of $X$ and $A_L$.
For example, the PBs of the variables $U$  are
$$\{U(x) \otimes  U(y)\}_{PB} = -{1\over \beta}\  U(x)\ t_a\otimes U(y)\ t_b\
K^{ab}(x,y).			\eqn\btwentythree$$
It is also possible to calculate the Poisson brackets of the variables  $U$ or
$V$ that lie in different representations of the group $G$.
To compare with the results of the authors of ref. [\chu], we note that they
use the variable $u(x)= U(x) M^{x\over l}$, where $x\in \bf{R}$. The PBs of
the $u$ variables can then be checked to be those given in this reference
\foot {We would like to thank Meifang Chu for pointing this out to us.}.

Because the symplectic form $\Omega$ factorises,
the PB of $X$ with $Y$ is zero, where $V=V(Y)$.  This  is consistent
with the Jacobi identities.  As a consequence of this, the PB of the current
$J_-(Y)$ with the current $J_+(X)$ is zero and the theory has two commuting
Kac-Moody algebras, one for each chiral sector.

To describe the WZW model we introduce a constraint $ Q = A_L - A_R$
on the phase space ($P_{LR}, \Omega$) of the LR model.  This constraint does
not generate other constraints and it is first class.  Moreover, under the PBs
of the LR model the constraint $Q$ generates the transformations of eqn. \aten.

The introduction of further gauge fixing conditions was
suggested in ref. [\chu ] as another way to construct the inverse
of the symplectic
form of the WZW model. However, it is not clear to us that the
PBs presented in this approach satisfy the Jacobi identities.

\section {The supersymmetric WZW model}

The Lagrangian of the (1,1) supersymmetric WZW  model is
$$ L= {k\over 4 \pi}\   (h_{ij}\     -
\  b_{ij})\  D_{+}\phi^i{} D_{-}\phi^j,			\eqn\none$$
where $h, b$ are defined as in the bosonic sigma model, $\phi$ is a map
from a (1,1) superspace with co-ordinates $\{z\}=\{x, t, \t^+, \t^-\}$ to a
group manifold $G$ and $D_-, D_+$ are superspace derivatives with $D_-^2= i
\partial_-$, $D_+^2= i \partial_+$ and $\{D_+,D_-\}=0$. The action
is invariant under the semilocal transformations $g(z) \rightarrow l(z^-) g(z)
r^{-1}(z^+)$ and the corresponding currents are
$$
J_-(z^-) = {k\over 4 \pi}\  g^{-1}\ D_-g, \quad
J_+(z^+) = -{k\over 4 \pi}\  D_+g\ g^{-1},			\eqn\nthree$$
where $\{z^+\}=\{x^+, \t^+\}$ and $\{z^-\}=\{x^-,\t^-\}$.

The equations of motion of the supersymmetric WZW model can be solved as in
the case of the bosonic WZW model.  The  general solution can be expressed in
the form
 $$ g(x,t,\t^+, \t^-) = U(z^+) M^{2t\over l}  V(z^-),     \eqn\nfour$$
where $g, U, V$ are periodic in the $x$ coordinate and the monodromy $M$ is on
a maximal torus of the group $G$.
The solution \nfour\ of the supersymmetric WZW model is invariant under the
analogoue of the right transformations given in eqn \aten, for the bosonic
model.

The symplectic current of the theory is given by
$S^{\pm}= \delta \phi^i \wedge \delta ({\d L / \d D_{\pm}\phi^i})$
and the symplectic form is
$$ \omega = -{i\over 2}\int^l_0 dx (D_+ S^- +D_- S^+ )|_{\t^{\pm}=0}.
\eqn\nseven$$

This symplectic form can be factorised into a left and a right symplectic
form by introducing different monodromies for the $U$ and $V$
sectors.  The left form $\omega_L$  written in a periodic basis
and in components is
$$\eqalign{
\omega_L =&- {k\over 8 \pi}\big{[} \int^{\l}_0 dx\ \{ tr(U^{-1} \delta
U\ \partial_x (U^{-1} \delta U) + {2{}i\over \l}(U^{-1} \delta U)^2\  A_L)\}
-i \delta_{ab} \delta\lambda^a_+ \delta\lambda^b_+
\cr &
+ {2\over
\l}\  tr (U^{-1} \delta U)_0\  \delta M_L M_L^{-1}\big{]} }    \eqn\neight$$
where $U, A_L$ are defined as in the bosonic case and $ - i\lambda^a_+ t_a =
(D_+U U^{-1})|_{\t^+=0}$  ($i\lambda^a_- t_a = (V^{-1} D_-V )|_{\t^-=0}$).
$U$ is periodic.

It is evident that $\omega_L$ factorises into  bosonic and fermionic sectors
and that these two sectors can be inverted separately.  The PBs of the
bosonic sector are the same as the PBs of the bosonic WZW model
and they are given in eqns \bfourteen\ - \bsixteen.  The remaining
PBs are
$$\eqalign{
\{ \lambda^a_+(x), \lambda^b_+(y)\}_{PB}&={i\over \beta} \delta^{ab}
\delta(x,y),  \cr
\{X^i(x), \lambda^a_+
(y)\}_{PB}&=\{A_L,\lambda_+^a(x)\}_{PB}=0.}           \eqn\nnine$$

These PBs satisfy the Jacobi identities.  Moreover, the Poisson
bracket algebra of the currents \nthree\ is the supersymmetric Kac-Moody
algebra.

Finally, we can introduce antiperiodic boundary conditions for the fermions
$\lambda_+$ and/or $\lambda_-$ instead of the periodic ones, due to the
existence of two spin structures on $S^1$.  In all these cases, the PBs of the
theory remain the same as in  eqns \bfourteen\ - \bsixteen\ and \nnine.

\section {The gauged WZW model}

Another class of interesting conformal field theories is the gauged WZW models.
We consider the case where one gauges a subgroup $H$ of the diagonal
subgroup of the $G\times G$ rigid symmetry group of the WZW model with target
space the group $G$.  The action of $H$ on $G$ is $ g\rightarrow hgh^{-1}$,
with $g\in G$ and $h\in H$.
The Lagrangian of the corresponding gauged WZW model can be written
$$ L= -{k\over 16 \pi}\   (h_{ij}\  \eta^{\mu \nu}\
\nabla_{\mu}\phi^i{} \nabla_{\nu}\phi^j + \  b_{ij}\  \epsilon^{\mu \nu}\
\partial_{\mu}\phi^i\  \partial_{\nu}\phi^j - \epsilon^{\mu \nu} a^A_{\mu}
\partial_{\nu}\phi^i w_{iA} + c_{AB} \epsilon^{\mu \nu} a^A_{\mu}\ a^B_{\nu}),
			\eqn\mtwo$$
where $\phi$ is a section of an $H$-bundle over $S^1\times \bf{R}$, $\nabla$ is
the covariant derivative of the connection $a$
($\nabla_{\mu}\phi^i=\partial_{\mu}\phi^i + a^A_{\mu} \xi_A$), $\xi_A$,
$A=1,\cdots, dim H$, are the vector fields generated by the group action of $H$
on $G$, $w$ is given by  $ \xi^i_A H_{ijk}= 3 \partial_{[j} w_{k]A}$ and
$c_{AB}= \xi^i_A w_{iB}$ (c.f. ref.[\osb]).

The equations of motion of the theory are
$$\eqalign {
\nabla_-( \nabla_+ g g^{-1}) + i F_{+-}(a) &=0
\cr
 (\nabla_+ g g^{-1})|_{LieH}=0,
\quad (g^{-1} \nabla_-g)|_{Lie H} &=0,}                          \eqn\mfour$$
where $\nabla_{\pm}g=\partial_{\pm} g +i a_{\pm} g - i g a_{\pm}$ and
$F(a)$ is the curvature of the connection $a$.    Equations \mfour\  imply that
$a$ is a flat connection. The general solution of the first equation of \mfour\
can then be written in the form
 $$ g(x,t) = m^{{t\over l}} U(x^+) M^{{2t \over l}} V(x^-) m^{{t\over l}}
						\eqn\mseven$$
where $g, U$ and $V$ are
periodic in the co-ordinate $x$ and $M$ ($M=exp iA$) is
the monodromy
familiar from the WZW model case.  $m$ is the holonomy of the flat
connection $a$.  We have used the $H$ gauge invariance to put $m$ on a maximal
torus of the group $H$.

The symplectic form of the gauged WZW model is given by
$$\eqalign{
 \omega (g, a) = - {k\over 8 \pi}\ \int^{\l}_0 dx\
tr\Big[
\delta g g^{-1} &\nabla_{-} (\delta g g^{-1}) + g^{-1} \delta g \nabla_+
(g^{-1} \delta g)  \cr
+2i \delta g g^{-1} & \delta a_- - 2i g^{-1} \delta g \delta a_+\Big].  }
			\eqn\sevena$$
Substituting the solution \mseven\ into this expression for $\omega (g, a)$, we
get the same symplectic form as in the case of the (ungauged) WZW model.
In particular  this $\omega$  does not depend on the connection $a$.
However, we
have not imposed the last two equations of motion of eqn. \mfour\ on
$U$ and $V$.
This turns out to be a rather involved problem which we will study
in a future publication.
Instead, we will examine a special case of the gauged
WZW model, that with gauge group $H=G$.  This is a topological model with
a finite number of degrees of freedom [\ed] and the phase space of the theory
is a finite dimensional manifold.

The space of solutions of the equations of motion \mfour\ of the topological
model on a cylinder is given by the space of flat connections $a$ of the group
$G$ times the space of constant elements $g$ that lie on a maximal torus $T$ of
the group $G$.  Thus the phase space of the topological model is $T\times T$.
The symplectic form (eqn. \sevena ) is
$$ \omega =  - {k\over 4 \pi}(\delta a\wedge \delta b)		\eqn\mnine$$
where $g=\exp i t_r b^r $, $r=1,\cdots, {\rm rank}G$.
The quantisation of quantum mechanical systems with phase space
structure similar to that of the topological model has been studied  in ref.
[\papa] using geometric quantisation methods.

\section {Concluding Remarks}

The topology and geometry of the spacetime is important in determining the
form of the PBs of any field theory and in particular the PBs of the WZW
model.  A change of geometry and topology of the spacetime has a
dramatic effect on the form of the PBs.  For example, consider the WZW
model on $\bf{R}^2$ with the flat Lorenzian metric.  In this case the model
has no monodromy and the operator that we have to invert is $D_{ab}={d\over
dx} \delta_{ab}$.  There are several difficulties in defining this theory
properly, one of them being  that the symplectic form $\omega$ is not well
defined for all $C^{\infty}$-functions $ u^{-1} \delta u $ and $ \delta v
v^{-1} $ on $\bf {R}$ and a careful treatment of the associated
analysis should
be undertaken. However, if we proceed formally, the operator $D$ has zero modes
and therefore does not have a unique inverse.  There is a family of inverses
with kernels
$$K_{ab} = \delta_{ab}\  \epsilon (x,y) + C_{ab}
					\eqn\done$$
where $\epsilon (x,y) = {1\over 2}, x>y$, $\epsilon (x,y) =
-{1\over 2}, y>x$ and $C$ is an undetermined constant antisymmetric matrix.
 The Jacobi identities impose an additional restriction on
$C$ which is that $C$ should satisfy the modified Yang-Baxter equation.  The
modified Yang-Baxter equation is  a Nijenhuis tensor
condition for the endomorphism $C$ of the Lie algebra $LieG$.
Solutions of this equation were discussed in ref.[\bab].

The phase space of a WZW model is an infinite dimensional space with
non-trivial topology.  In general, there is no  unique way to quantise
phase spaces with  non-trivial topological and global structure [\isham].
One reason for this  is that there is no  \lq\lq natural" choice of a set of
variables with respect to which we can quantise the system.  Different choices
of variables may lead to inequivalent quantisations.   One set of variables
that has already been used to quantise the WZW models is the set of currents
$\{J_+, J_-\}$ [\wit].  However there may be other choices of variables which
may give new insight into the quantum structure of the model.  For example, if
this theory is quantised in terms of the currents the monodromy does not play
any role in the quantum theory of the WZW model.

ACKNOWLEDGEMENTS

We would like to thank C. M. Hull and P.S. Howe for discussions.  This work was
supported by the SERC.

\refout

\end